\documentclass[journal=jpclcd,manuscript=letter,layout=twocolumn]{achemso}
\usepackage{dcolumn}
\usepackage{amssymb,amsmath}
\usepackage{graphicx}
\usepackage{hhline}
\usepackage{textcomp}
\usepackage{ADDmacros_updated}

\author{Vikram Plomp}
\author{Xu-Dong Wang}
\affiliation{Radboud University, Institute for Molecules and Materials, Heijendaalseweg 135, 6525 AJ 
Nijmegen, the Netherlands}
\author{Jacek K{\l}os}
\affiliation{University of Maryland, Department of Physics, Joint Quantum Institute, 
College Park, MD, United States of America}
\author{Paul J. Dagdigian}
\affiliation{Johns Hopkins University, Department of Chemistry, Baltimore, MD, United States of America}
\author{Fran{\c c}ois Lique}
\affiliation{Universit{\'e} de Rennes, Institut de Physique de Rennes, 263 avenue du G{\'e}n{\'e}ral 
Leclerc, 35042 Rennes CEDEX, France}
\email{francois.lique@univ-rennes1.fr}
\author{Jolijn Onvlee}
\author{Sebastiaan Y.T. van de Meerakker}
\email{basvdm@science.ru.nl}
\affiliation{Radboud University, Institute for Molecules and Materials, Heijendaalseweg 135, 6525 AJ 
Nijmegen, the Netherlands}

\title{Imaging Resonance Effects in C + H$_2$ Collisions using a Zeeman Decelerator}
 
\begin{tocentry}
	{\includegraphics[width=5cm]{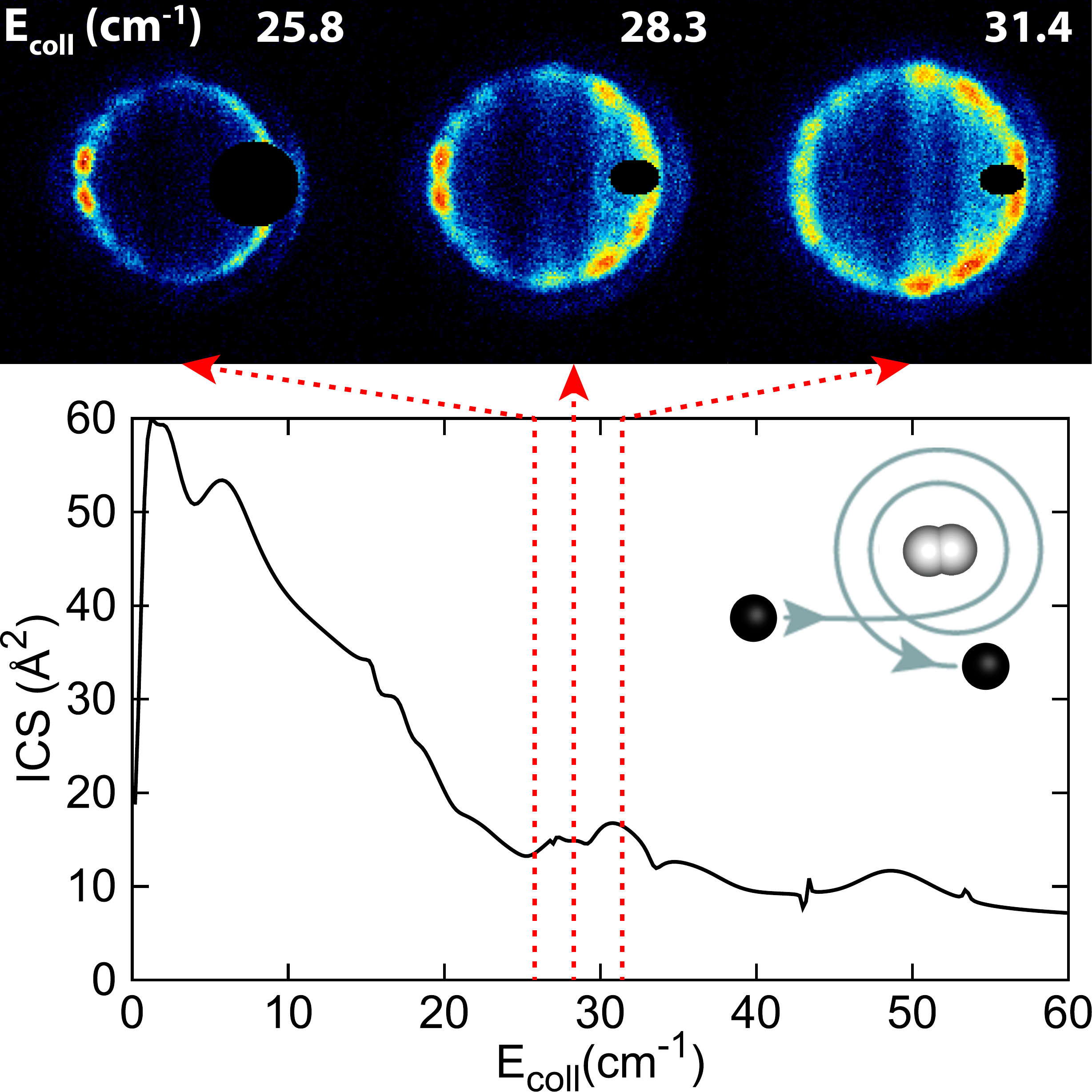}}
	\\\\
	The used combination of Zeeman deceleration, near-threshold ionization detection and velocity map imaging allowed for high-resolution imaging of C-atom scattering distributions after interaction with H$_2$. The observed rapid changes in the distributions as a function of collision energy ($E_{coll}$) can be attributed to the effects of scattering resonances.
\end{tocentry}

\begin{document}

\begin{abstract}
	An intriguing phenomenon in molecular collisions is the occurrence of scattering resonances, which originate from bound and quasi-bound states supported by the interaction potential at low collision energies. The resonance effects in the scattering behaviour are extraordinarily sensitive to the interaction potential, and their observation provides one of the most stringent tests for theoretical models. We present high-resolution measurements of state-resolved angular scattering distributions for inelastic collisions between Zeeman-decelerated C($^3P_1$) atoms and \textit{para}-H$_2$ molecules at collision energies ranging from 77~cm$^{-1}$ down to 0.5~cm$^{-1}$. Rapid variations in the angular distributions were observed that can be attributed to the consecutive reduction of contributing partial waves and effects of scattering resonances. The measurements showed excellent agreement with distributions predicted by \textit{ab initio} quantum scattering calculations. However, discrepancies were found at specific collision energies, which most likely originate from an incorrectly predicted quasi-bound state. These observations provide exciting prospects for further high-precision and low-energy investigations of scattering processes that involve paramagnetic species.
\end{abstract}

To acquire a detailed understanding of molecular interactions has been an important goal in physical chemistry for decades \cite{LevineBernstein1987}, and many refined experiments have been designed to test existing theoretical models and support their further development \cite{Yang:ModTrendChemReactDyn, TutorialsMRD, Casavecchia2000, Kohguchi2002, Picard2019}. The potential energy surfaces (PESs) underlying the molecular interactions can be probed through measurements that reflect the integral (ICSs) or differential (DCSs) cross sections, and both have been provided by crossed-beam experiments that studied molecular collisions in the gas phase \cite{Liu2006, Yang2011, Naulin2014, Aoiz2015}. In this regard, the high-resolution measurements of angular scattering distributions afforded by the combination of Stark deceleration to control collision partners and velocity map imaging (VMI) to probe collision products has proved especially powerful to study DCSs \cite{Onvlee2014,Onvlee:CPC17:3583}. It enabled the observation of delicate features and allowed to gain new insights into scattering mechanisms. Recent examples include the direct imaging of quantum diffraction oscillations \cite{VonZastrow2014,Jongh:JCP:2017,Onvlee:nchem9:226}, the measurement of correlated excitations in bimolecular collisions \cite{gao2018observation,Besemer:nchem14:664,Tang:Science:379}, and the probing of scattering resonances at low collision energies \cite{vogels2015imaging,vogels2018scattering,Jongh:Science:2020,Jongh:nchem:2022}.

However, Stark deceleration is restricted to molecules that have an electric dipole moment, and high-resolution VMI measurements require efficient near-threshold resonance-enhanced multiphoton ionization (REMPI) schemes that are sparsely available. Thus, the number of systems for which the full potential of this approach could be exploited remained limited. The recent demonstration of Zeeman deceleration in crossed beam scattering experiments employing VMI \cite{Plomp2020,Plomp:JPCA:2023}, and in particular the combination with recoil-free vacuum ultraviolet (VUV) based REMPI detection \cite{Plomp:JPCL:2021}, has alleviated these restrictions.

While the experimental observation of diffraction oscillations provided an exquisite demonstration of the approach and a stringent test for theoretical descriptions \cite{Plomp2020,Plomp:JPCL:2021}, other intriguing quantum effects occur that provide an even more sensitive test for interaction models. At low collision energies, bound and quasi-bound states supported by the interaction potential give rise to scattering resonances that can cause rapid and dramatic changes in both the ICSs and DCSs. As the energy of these states and their effect on the scattering behaviour are extraordinarily sensitive to the shape of the potential over the full range of interaction, the observation of these resonance effects can be considered one of the most stringent tests for theoretical models. Thus, there has been a continuous interest in performing controlled collision experiments at low energies \cite{softley2009,REVStuhl2014,ReviewToscano2020,SoftleyHeazlewood2021,ReviewWu2022,Softley:2023}.

On the theoretical side, a particular challenge lies in the quantum mechanical modelling of non-adiabatic coupling effects between multiple interaction potentials. These arise in inelastic collisions of open-shell species with other atoms or molecules, and result in a breakdown of the Born-Oppenheimer approximation. Typical examples are the spin-orbit (de-)excitation of ground-state atomic carbon, C($^3P_j$) $\rightarrow$ C($^3P_{j'}$), in collisions with He or H$_2$ \cite{Bergeat2018,Bergeat2019,Klos2018,Picard2002,Monteiro1987,Lavendy1991,Staemmler1991}. Here, the triatomic and reactive nature of the interaction of C with H$_2$ provides an additional challenge for theoretical descriptions \cite{Klos2018,Bergeat2019}.

The given examples are of specific importance to astrochemistry as He and  H$_2$ are the most abundant collision partners in space, while C is the fourth most abundant element \cite{Wolfire1995}. Atomic carbon in its electronic ground state is especially abundant in many interstellar regions ranging from molecular clouds to planetary nebulae \cite{Zmuidzinas1988,Bensby2006} and plays a crucial role in interstellar molecular chemistry and synthesis of many carbon-rich molecules \cite{Henning1998,Hollenbach1999}. The equilibrium between radiative processes and collision induced fine-structure transitions of atomic carbon is specifically important in interstellar cloud cooling and provides a probe for astrophysical conditions in these regions \cite{Wolfire1995,Neufeld1995,Bensch2003,Tanaka2011}.

At temperatures below $\sim100$~K scattering resonances with a profound influence on the ICSs have been predicted and observed for collisions of C($^3P_0$) with both He and H$_2$/D$_2$ \cite{Bergeat2018,Klos2018,Bergeat2019}. The resonance features observed in the ICSs of C($^3P_0$) + He collisions were in excellent agreement with theoretical predictions and allowed for a detailed description \cite{Bergeat2018}. However, for the C($^3P_0$) + H$_2$/D$_2$ collisions the resonance spectrum is rather congested and, despite the good agreement with theory, this hampered the observation and assignment of individual features in the experimental ICSs \cite{Klos2018,Bergeat2019}.

The use of a Zeeman decelerator to prepare velocity-controlled packets of carbon atoms with narrow velocity and angular spreads can enable an improved comparison with theory. More importantly, the combination of Zeeman deceleration, VMI and recoil-free VUV-based detection of carbon atoms, as demonstrated recently \cite{Plomp:JPCL:2021}, allows for high-resolution experimental investigations of quantum-state-resolved DCSs for these processes. Such investigations could provide even more stringent tests for theory as well as further insight into the underlying scattering mechanisms.

In this work, we experimentally probed state-resolved DCSs for the spin-orbit de-excitation collision process C($^3P_1$) + \textit{p}-H$_2$ ($j=0$) $\rightarrow$ C($^3P_0$) + \textit{p}-H$_2$ ($j=0$) with high resolution through a crossed beam experiment employing a Zeeman decelerator, VUV-based REMPI detection and VMI. The C atom is well suited for manipulation using magnetic fields \cite{Jankunas2015, Karpov2020}, and the Zeeman decelerator thus allowed the preparation of velocity-controlled packets of C($^3P_1$) atoms with narrow velocity and angular spreads. The combination of VUV-based REMPI detection and VMI allowed to efficiently image the velocity distribution of scattered carbon atoms without ion-recoil \cite{Plomp:JPCL:2021}. Further details on the production and characterization of the prepared packets of C($^3P_1$) as well as details on the employed detection methods have been presented elsewhere \cite{Plomp:JPCL:2021}. To cover a large range of collision energies two different experimental geometries were implemented. For intermediate collision energies ($E_{coll}=$~28 to 77 cm$^{-1}$) a setup with a $\degree{46}$ beam intersection angle was employed. Collision energies down to $0.5 \text{ cm}^{-1}$, required to probe the onset of the resonance regime, were obtained by adding an extension to the Zeeman decelerator that allows for a small beam intersection angle of $\degree{4}$. The exceptional resolution of the experiment allowed us to fully resolve diffraction oscillations over a large range of collision energies. Rapid changes in the scattering distributions were observed that are attributed to the effects of scattering resonances as well as the consecutive reduction of contributing partial waves as the collision energy decreases. In general, excellent agreement was found with simulations based on \textit{ab initio} calculations of the involved PESs. However, some distinct discrepancies were found at collision energies between 39 and 50 cm$^{-1}$. A more detailed investigation revealed that these discrepancies most likely arise from a quasi-bound state that is erroneously predicted to exist by the employed theoretical models. As this quasi-bound state can already cease to exist with a 0.5~cm$^{-1}$ reduction of the centrifugal barrier, this can be attributed to a very minor inaccuracy in the theoretical modelling of the PESs. These results provide a testimony of the precision of the employed approach, and underline the prospects to study a large range of collision processes with an unprecedented level of precision.

\begin{figure*}[!htb]
   \centering
    \resizebox{1.0\linewidth}{!}
    {\includegraphics{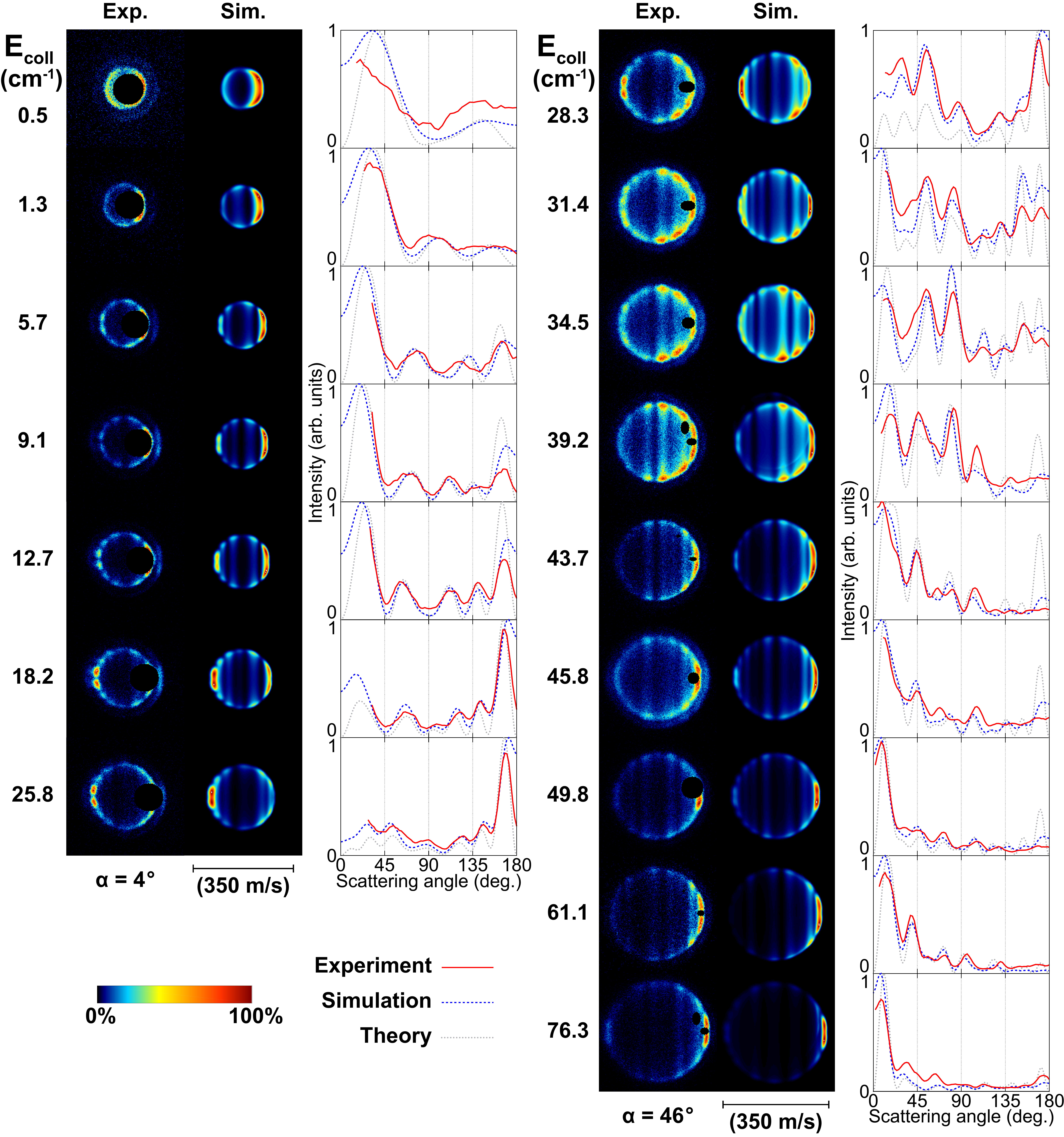}}
    \caption{Experimental (Exp.) and simulated (Sim.) scattering images for the  C($^3P_1$) + \textit{p}-H$_2$ ($j=0$) $\rightarrow$ C($^3P_0$) + \textit{p}-H$_2$ ($j=0$) process ($16.4$~cm$^{-1}$ de-excitation) at selected collision energies ($E_{coll}$). The faint outer rings in the experimental images arise from co-decelerated C($^3P_2$) ($43.4$~cm$^{-1}$ de-excitation). The extracted angular scattering distributions are shown beside the images for both experiment (red solid lines) and simulation (blue dashed lines), together with the DCSs from theory (gray dotted lines). A beam intersection angle of $\alpha = \degree{4}$ (left panels) or $\alpha = \degree{46}$ (right panels) was employed.}
    \label{fig:CH2data}
\end{figure*}

The scattering images that were recorded for the spin-orbit de-excitation process C($^3P_1$) + \textit{p}-H$_2$ ($j=0$) $\rightarrow$ C($^3P_0$) + \textit{p}-H$_2$ ($j=0$) at selected mean collision energies ($E_\text{coll}$) are depicted in \autoref{fig:CH2data}. The images are presented such that the relative velocity vector is directed horizontally, with forward scattering angles positioned at the right side. Small segments of the images are masked where the initial beam gives a contribution to the signal. Each pixel corresponds to a velocity of about 2.5 m/s. It is apparent from the images that the scattering distribution varies rapidly with the collision energy. The observed changes can be further quantified through the angular scattering distributions extracted from the experimental image intensities within a narrow annulus around the observed scattering rings (also depicted in \autoref{fig:CH2data}). Besides very pronounced changes like the sudden appearance or disappearance of diffraction peaks, this also reveals more subtle changes like shifts in the peak positions or changes in their relative intensities.

The extracted angular scattering distributions can be directly compared to the distributions obtained from simulated images (see \autoref{fig:CH2data}). Our image simulations are based on theoretical predictions obtained from \textit{ab initio} quantum mechanical close-coupling (QM CC) calculations \cite{Alexander1998}, in combination with particle trajectory simulations on our Zeeman decelerator apparatus. The QM CC calculations employ the state-of-the-art C($^3P_j$) + \textit{p}-H$_2$ PESs of K{\l}os \textit{et al.} \cite{Klos2018,Bergeat2019} computed using the explicitly correlated multireference configuration interaction method (ic-MRCI-F12) \cite{Shiozaki:11} with a large atomic basis set (vide infra).  The simulated images are shown alongside the experimental ones, and are analyzed analogously to their experimental counterparts to acquire predicted angular scattering distributions that take into account the spatial, temporal and velocity spreads of the used atomic and molecular beams, as well as kinematic effects on the scattering distributions \cite{VonZastrow2014,Plomp2020}. Generally, an excellent agreement is observed between the experimental and simulated scattering distributions. The observed changes in the peak positions and relative intensities are qualitatively reproduced.

At collision energies below 35~cm$^{-1}$ and above 55~cm$^{-1}$ there is not only excellent qualitative agreement between the experiments and simulations, but also a very good quantitative agreement. Here, both the positions and intensities of the diffraction peaks in the experimental scattering distributions are accurately reproduced by the simulations. However, at intermediate collision energies a clear discrepancy can be observed between the experimental and simulated distributions. Around 44~cm$^{-1}$ the theory predicts a strong peak in the DCS in the backward scattering direction (close to 180{\textdegree} scattering angle), which gives rise to a smaller but distinct peak in the backward direction for the simulated scattering distributions. This peak can be observed in the simulated images and corresponding angular scattering distributions for a relatively broad range of energies, i.e. between 39 and 50~cm$^{-1}$. While some backscattering is also observed in the experimental images and scattering distributions for this energy range, a distinct peak in the backward direction can not be found there.

\begin{figure}[!b]
    \begin{center}
    \resizebox{1.0\linewidth}{!}
    {\includegraphics{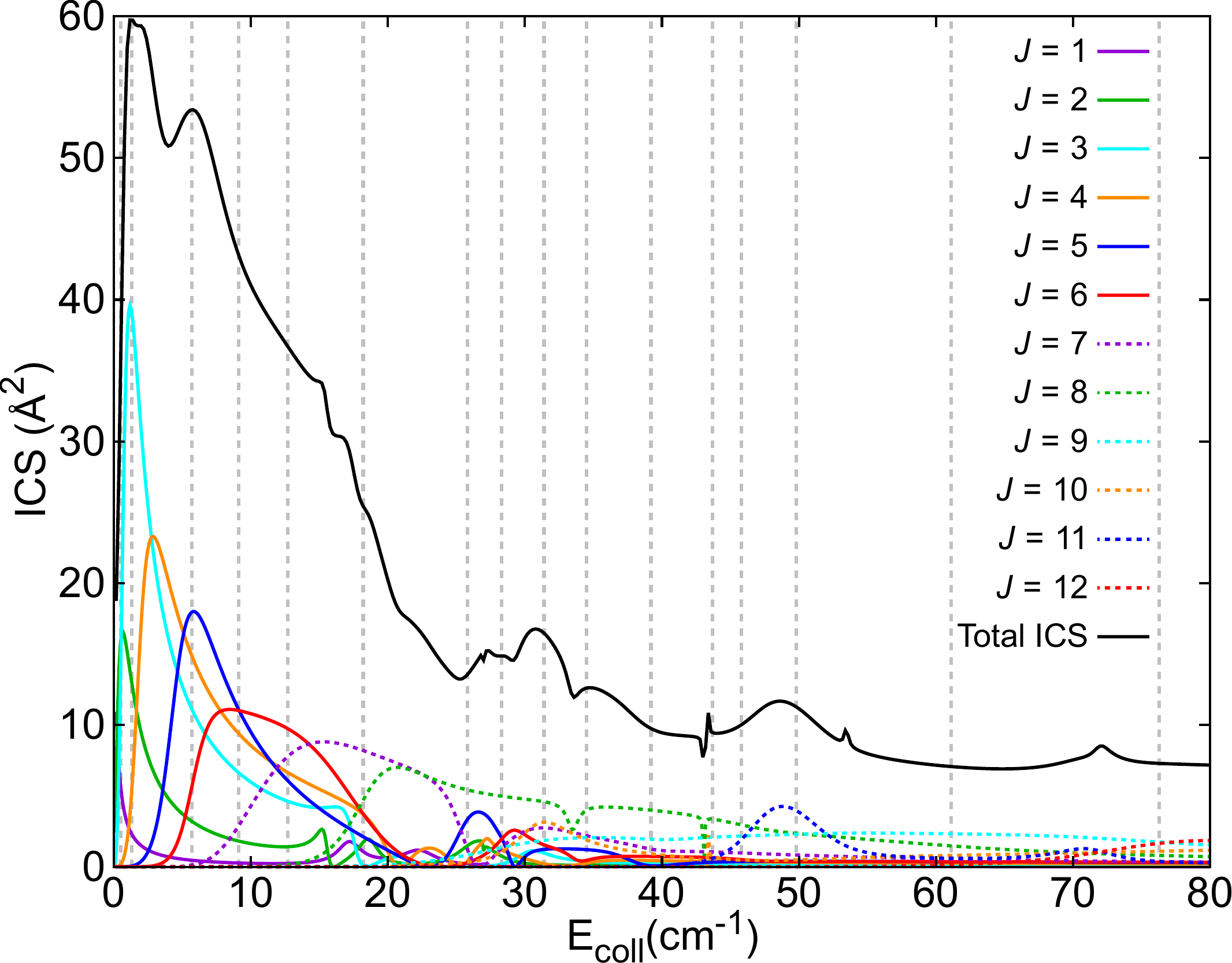}}
    \caption{Theoretical predictions for the partial contribution of each of the relevant total angular momentum states (${J} = 1$ to ${J} = 12$) to the total ICS of the C($^3P_1$) + \textit{p}-H$_2$ ($j=0$) $\rightarrow$ C($^3P_0$) + \textit{p}-H$_2$ ($j=0$) spin-orbit de-excitation process (black line). The selected mean collision energies probed in the experiment are indicated with vertical dashed lines.}
    \label{fig:CH2ICS}
    \end{center}
\end{figure}

The observed rapid variations of the scattering distribution with collision energy are a strong indication of the occurrence of resonance effects. When the collision energy becomes resonant with a quasi-bound state supported by the interaction potential, the scattering contribution of specific partial waves is altered, causing a sudden change in the DCS that reflects their interference pattern. Thus, the measurements of angular scattering distributions provide a fingerprint of the partial-wave composition of the scattering process \cite{Onvlee2016,vogels2018scattering,Jongh:Science:2020}. The predicted contribution of partial waves corresponding to specific total angular momenta ($J$) of the collision complex can be visualized through their partial ICSs, see \autoref{fig:CH2ICS}. The de-excitation process studied here is accurately described using only $J = 1$\footnote[1]{$J = 0$ does not contribute to this process, as the corresponding channel violates conservation of total parity.} to $J = 12$ for collision energies up to $\sim 70 \text{ cm}^{-1}$. At higher energies additional partial waves (with $J > 12$) make a significant contribution. It can be seen that the contribution of individual partial waves indeed changes rapidly over this energy range, with distinct peaks and valleys that correspond to scattering resonances. The rapid variations in the angular scattering distributions observed in the experiments are therefore interpreted in terms of the occurrence of scattering resonances, as well as the consecutive addition of contributing partial waves as the energy increases.

From the partial ICSs (see \autoref{fig:CH2ICS}) it can also be seen that at the energies of the apparent mismatch between experimental and predicted scattering distributions, around 44~cm$^{-1}$, the partial wave corresponding to ${J} = 11$ starts to contribute to the collision process. To further investigate the cause of the discrepancies we employed the adiabatic bender model \cite{Holmgren:1977,Alexander:1994} to analyze the underlying resonances. The two adiabatic bender curves for the ${J} = 11$ total angular momentum states corresponding to the C($^3P_1$) initial and C($^3P_0$) final state of the studied de-excitation process are depicted in \autoref[(a)]{fig:J1to10}. The energies of the van der Waals stretch levels supported by the curves were derived using a discrete variable representation method \cite{Colbert:1992,Tao:2007}. We identified a quasi-bound state at a total energy of about 59.9~cm$^{-1}$, corresponding to 43.5 cm$^{-1}$ collision energy after subtraction of the $16.4$~cm$^{-1}$ de-excitation energy. The average C-H$_2$ inter-particle distance for this state is $\left<R\right>=11.86$~bohr, while neighbouring bound states have $\left<R\right>\approx20$~bohr. This quasi-bound state is anticipated to strongly enhance the contribution of the ${J} = 11$ partial wave around this collision energy. It is thus expected to give rise to the broad resonance feature observed in the ${J} = 11$ partial ICS from about 40 to 55 cm$^{-1}$ and consequently induce the predicted DCS peak in the backward scattering direction. However, this quasi-bound state can already cease to exist when the centrifugal barrier is reduced by only about 0.5~cm$^{-1}$. Thus, a very minor change in the PESs can drastically change the contribution of the ${J} = 11$ partial wave around the 43.5~cm$^{-1}$ collision energy.

\begin{figure*}[!htb]
   \centering
    \resizebox{0.9\linewidth}{!}
    {\includegraphics{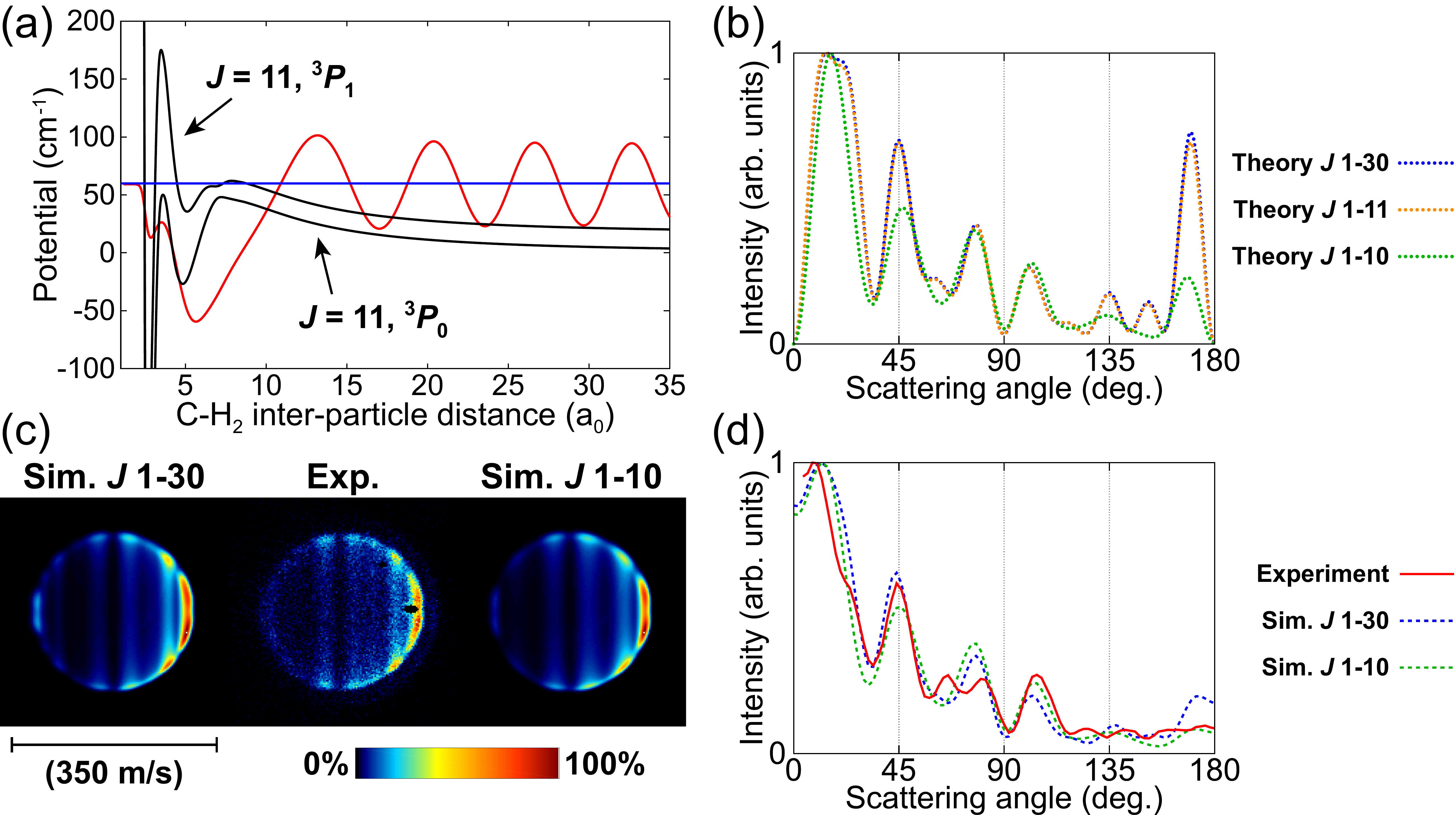}}
    \caption{(a) Adiabatic bender curves (black lines) for the ${J} = 11$ total angular momentum states corresponding to the C($^3P_1$) initial and C($^3P_0$) final state of the C($^3P_1$) + \textit{p}-H$_2$ ($j=0$) $\rightarrow$ C($^3P_0$) + \textit{p}-H$_2$ ($j=0$) spin-orbit de-excitation process. The total energy of a predicted quasi-bound state (59.9 cm$^{-1}$) is indicated by the blue line and the corresponding wave function is illustrated by the red line. (b) Theoretical DCSs at $E_{coll}=43.7$~cm$^{-1}$ obtained when considering different ranges of total angular momentum states: ${J} = 1$~to~30 (blue line), ${J} = 1$~to~11 (orange line) and ${J} = 1$~to~10 (green line). (c) Experimental (Exp.) scattering image at $E_{coll}=43.7$~cm$^{-1}$ and simulated (Sim.) images obtained using theoretical DCSs that include ${J} = 1$~to~30 or ${J} = 1$~to~10. The extracted angular scattering distributions are depicted in (d).}
    \label{fig:J1to10}
\end{figure*}

To illustrate the influence of the ${J} = 11$ partial wave on the scattering behaviour at the experimental mean collision energy of 43.7~cm$^{-1}$, theoretical DCSs were calculated that consider different ranges of partial waves, see \autoref[(b)]{fig:J1to10}. The theoretical prediction that includes a large range of partial waves (${J} = 1$~to~30) is nearly identical to the one that considers only ${J} = 1$~to~11, indicating that at this collision energy the calculations already converge when taking only the first eleven total angular momentum states into account. However, when ${J} = 11$ is also excluded the theoretical DCS changes significantly and the intensity of the peak in the backward scattering direction is strongly reduced. Corresponding image simulations at $E_{coll} =$~43.7~cm$^{-1}$, based on the theoretical DCSs that include either ${J} = 1$~to~30 or ${J} = 1$~to~10, are presented in \autoref[(c)]{fig:J1to10} and are compared to the experimentally obtained image at this energy. The angular scattering distributions extracted from these images are depicted in \autoref[(d)]{fig:J1to10}. It can be seen in both the images and the corresponding angular distributions that the predicted backward scattering is significantly reduced when the contribution of ${J} = 11$ is neglected. Excluding ${J} = 11$ in the theoretical calculations thus results in a better agreement with the experimental results at this collision energy, especially for the backward scattering direction. It should be noted, however, that the contribution of ${J} = 11$ does not solely arise from the identified quasi-bound state and full neglect of the corresponding partial wave can not give a fully accurate description of the experiments.

Based on these investigations, we can conclude that the mismatch between experimental and predicted scattering distributions in the range of $E_{coll}=$~39 to 50 cm$^{-1}$ is most likely caused by a quasi-bound state that is erroneously predicted to exist by the employed theoretical models, and that gives rise to an enhanced contribution of the ${J} = 11$ partial wave around these energies. As the quasi-bound state and corresponding resonance can already disappear with a 0.5~cm$^{-1}$ reduction in the centrifugal barrier, this can be attributed to a very minor inaccuracy in the theoretical modelling of the PESs. This illustrates that the experiments demonstrated in the current work provide an extraordinarily sensitive probe for theoretical models, and can be used to test and improve existing theoretical descriptions as well as our understanding of scattering processes. Finally, it should be noted that our findings could also help to further explain the previously observed discrepancies between theoretical predictions and measurements of ICSs for excitation collisions between C($^3P_0$) and H$_2$/D$_2$ \cite{Klos2018,Bergeat2019}.

The experimental results presented here are the first demonstration of the observation of scattering resonance effects in crossed-beam experiments employing a Zeeman decelerator, and underline the prospects for further high-precision and low-energy investigations. The sensitivity and attained collision energy range are similar to those in recent experiments employing Stark decelerators, which have proved to probe theoretical models with exceptional accuracy, and for example enabled detailed characterizations of scattering resonances \cite{Jongh:Science:2020,Jongh:nchem:2022} as well as revealing novel scattering mechanisms in bimolecular collisions \cite{Besemer:nchem14:664,Tang:Science:379}. While the technique of Stark deceleration is limited to molecules that have an electric dipole moment, the method of Zeeman deceleration utilized here can be applied to a large class of paramagnetic species \cite{Meerakker:CR112:4828,ReviewWu2022}. Together with the use of VUV light for REMPI detection, this opens up new perspectives for detailed experimental investigations on a variety of scattering processes. For example, the recently reported near-threshold VUV REMPI schemes for H/D \cite{Yuan2018} or O($^3P$) atoms \cite{Wang2021} provide the interesting opportunity to investigate elementary reactive scattering processes, such as C + O${_2}$ $\rightarrow$ CO + O\cite{Geppert2000,Karpov2020,Veliz2021} or complex-forming reactions between Zeeman decelerated atoms and H$_2$ molecules \cite{Aoiz2006, Guo2012, Yang2007}. A promising candidate is the S($^1D$) + H${_2} \rightarrow$ SH + H reaction, for which distinct resonance features have been predicted to occur at low energies in both ICSs and DCSs \cite{Jambrina2021}. Although a large body of work exists on these types of reactive systems \cite{Liu2006,Yang2007,Yang2011,Naulin2014,AnnuRevZhang2016,ReviewAshfoldXueming2018,PerspectiveLi2020}, there has been a continuous interest to study their scattering behaviour under controlled and low-energy conditions \cite{softley2009,REVStuhl2014,ReviewToscano2020,SoftleyHeazlewood2021,ReviewWu2022,Softley:2023} such as those provided by our combination of techniques.

\section*{Experimental methods}
\begin{figure}[!htb]
   \centering
    \resizebox{1.0\linewidth}{!}
    {\includegraphics{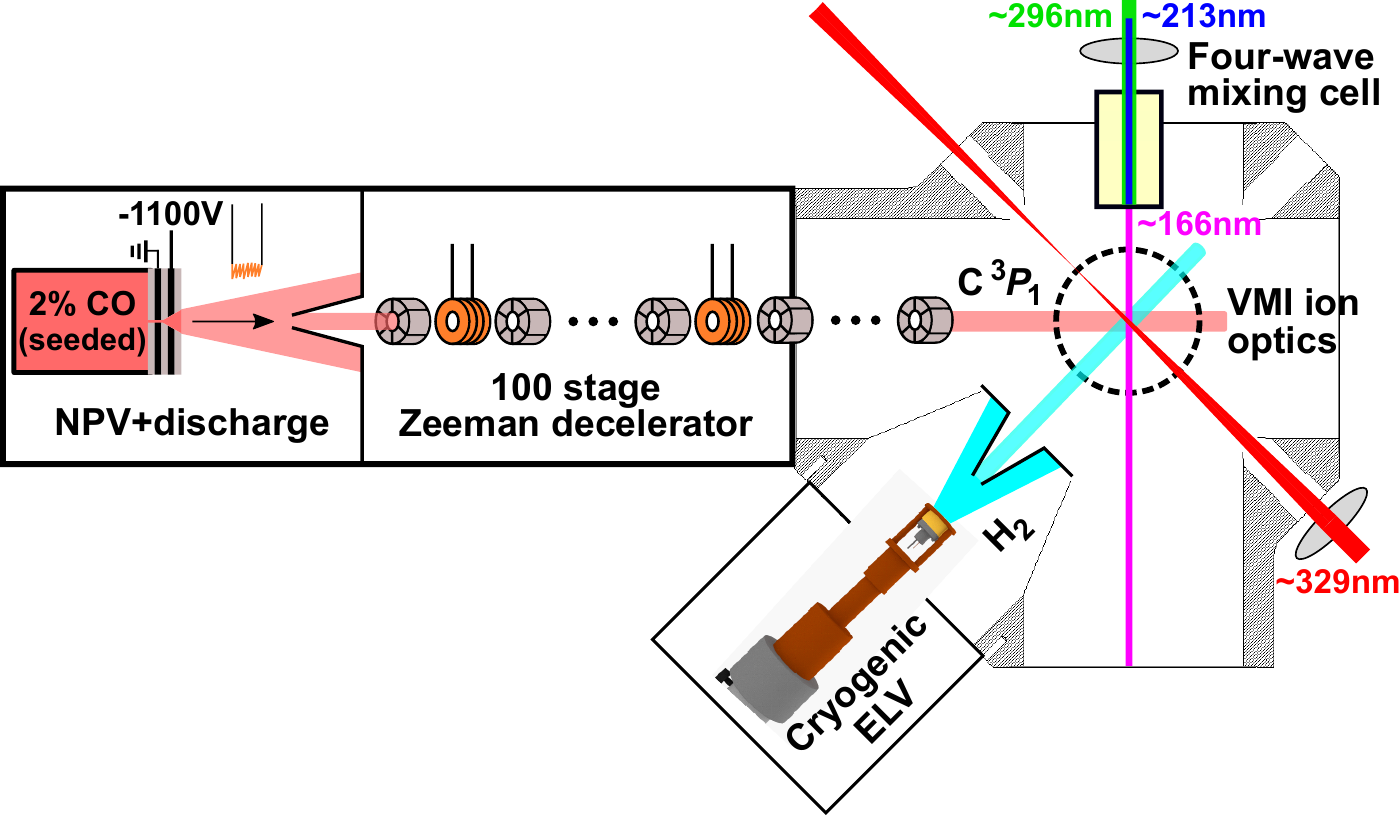}}
    \caption{Schematic depiction of the crossed-beam setup employing a $\degree{46}$ crossing angle. A decelerator extension (not depicted here) that allowed for a $\degree{4}$ crossing angle was used to obtain lower collision energies (see text). The image was adapted from reference \cite{Plomp:JPCL:2021} with permission from the authors.}
    \label{fig:Setup}
\end{figure}

A beam of carbon atoms, C($^3P_j$), was generated by running an electric discharge through an expansion of 2\% CO seeded in noble gas (see \autoref{fig:Setup}), using a Nijmegen pulsed valve (NPV) with discharge assembly \cite{Ploenes2016}. Mixtures of Kr, Ar, Ne and He were employed as seed gas to cover a large range of initial velocities. After the expansion, the majority of the carbon atoms resided in the $^3P_0$ ground state spin-orbit level, while the $^3P_{1,2}$ levels were much less populated. This beam of carbon atoms then passed a skimmer and entered the Zeeman decelerator, of which a detailed description is given elsewhere \cite{Cremers2019}. Briefly, it consists of an alternating array of pulsed solenoids and permanent magnetic hexapoles that allow independent control over the longitudinal and transverse motion of paramagnetic species, respectively. The decelerator contains a total of 100 solenoids and 101 hexapoles, and was operated at a repetition rate of 20 Hz. The C-atom $^3P_1$ state has a magnetic moment of 1.5 $\mu_B$ and atoms in this state can be effectively manipulated with the decelerator. Although atoms in the C($^3P_2$) state, with a magnetic moment of 3 $\mu_B$, were co-decelerated with the C($^3P_1$) atoms, their density in the beam was significantly lower. While the $^3P_0$ state had a much higher initial population, it is almost insensitive to magnetic fields. The corresponding free flight through the decelerator reduced the final $^3P_0$ atom contribution to negligible levels. Thus, the decelerator was used to obtain packets of mainly C($^3P_1$) with controlled mean velocity ($v_C=$~300 to 1450 m/s), and narrow velocity and angular spreads. Further details on the production and characterization of the prepared packets of C($^3P_1$) can be found elsewhere \cite{Plomp:JPCL:2021}. A series of additional hexapoles guided the packets of C($^3P_1$) towards the interaction region where they were intersected by a beam of H$_2$ molecules. The neat H$_2$ beam was produced using an Even-Lavie valve (ELV) that was cryogenically cooled (25~-~70 K) to control the mean velocity ($v_{H_2}=$~870 to 1320 m/s). It is noted that H$_2$ co-exists in two forms, \textit{ortho}- and \textit{para}-hydrogen, for which a significantly different scattering behaviour is predicted \cite{Klos2018}. A pure ($\gtrsim98\%$) beam of \textit{p}-H$_2$ ($j=0$) was obtained by repeated liquefaction and subsequent evaporation of normal H$_2$ gas in the presence of nickel(II)-sulfate catalyst before expansion \cite{Shuai2020}.

The measurements with a mean collision energy ($E_{coll}$) between 28 and 77~cm$^{-1}$ were performed using a setup where the packets of C($^3P_1$) exiting the decelerator are guided to the interaction region by 13 additional hexapoles, and are intersected by the H$_2$ beam at an angle of $\alpha=\degree{46}$ about 368.5 mm from the decelerator exit. The measurements for $E_{coll}=$~0.5 to 26~cm$^{-1}$ were performed by employing an extension to the Zeeman decelerator instead, that allows for a small beam intersection angle of $\alpha=\degree{4}$. The extension features a 33-hexapole guide as the interaction region lies about 871.5 mm from the decelerator exit here. While the extension also houses an additional 27 coils to maintain control over the C($^3P_1$) packets in the longitudinal direction, these were not employed in the experiments reported here. After scattering, the product C($^3P_0$) atoms were state-selectively ionized using a near-threshold (1+1$'$) (VUV+UV) resonance-enhanced multiphoton ionization scheme \cite{Plomp:JPCL:2021}, and detected with the use of high-resolution VMI ion optics that allows for accurate mapping of large ionization volumes \cite{Plomp2021}. Due to the narrow velocity spreads of the decelerated C atoms the scattering signal arising from the contribution of co-decelerated initial C($^3P_2$) could be well separated from the main C($^3P_1$) contribution.

Calibration of the VMI system was conducted through measurements on the C($^3P_1$) packets exiting the decelerator at different velocities, resulting in a velocity conversion factor of about 2.41 and 2.55 m/s per pixel for the high- and low-energy setup, respectively. The beam intersection angles were calibrated through VMI measurements on Kr atoms exiting both valves at several velocities, detected in the interaction region using a sensitive 2+1$'$ (212~nm + 326~nm) REMPI-scheme. The results were $\alpha=\degree{45.95} \pm \degree{0.08}$ and $\alpha=\degree{4.02} \pm \degree{0.14}$ ($95\%$ confidence interval) for the high- and low-energy setup, respectively. However, the kinematics of the experiment in the $\degree{4.02}$ setup result in an effective intersection angle of $\alpha'=\degree{4.13}$, as determined through particle trajectory and collision simulations. This effective angle was used for the image simulations and calculation of the collision energies. The mean H$_2$ velocities were calibrated by determination of the angle $\beta$ between the axis of the Zeeman-decelerated beam and the relative velocity vector in a (calibration) scattering image. Together with $v_C$ and $\alpha$ (or $\alpha'$), this angle $\beta$ enables the calculation of $v_{H_2}$ through trigonometric rules. The expected error of less than 10~m/s allows for an accurate calculation of $E_{coll}$. The effective longitudinal velocity spread of the H$_2$ beam was around $5\%$ of $v_{H_2}$ (full width at half maximum), as obtained through comparison of simulated image blurring with experimental observations. The collision energy spreads were determined through the image simulations that take into account the spreads of the beams as well as the kinematics of the experiment.

\section*{Theoretical methods}
For the calculation of the integral and differential cross sections for the collisional process C($^3P_j$) + \textit{p}-H$_2$ ($j=0$) $\rightarrow$ C($^3P_{j'}$) + \textit{p}-H$_2$ ($j=0$) we employed the exact quantum mechanical close-coupling (QM CC) approach \cite{Alexander1998}, which allows the study of collisions without any approximations. The calculations were performed using the \textit{ab initio} highly-correlated C($^3P_j$) + \textit{p}-H$_2$ PESs of K{\l}os \textit{et al.} \cite{Klos2018,Bergeat2019}. Briefly, the interaction of the open-shell C($^3P$) atom with the H$_2$($^1\Sigma_g^+$) molecule gives rise to three adiabatic PESs: $1^3A''$, $2^3A''$ and $1^3A'$. Two are of $A''$ symmetry (wave function anti-symmetric with respect to reflection in the triatomic plane) and one is of $A'$ symmetry (symmetric with respect to reflection in the triatomic plane). The two adiabats of the same $A''$ symmetry will avoid crossing with each other, and for the full description of the dynamics of this system an off-diagonal diabatic PES needs to be calculated. The PESs were computed using the internally contracted, explicitly correlated variant of the multireference configuration interaction method (ic-MRCI-F12) \cite{Shiozaki:11} that employed all-electron correlation-consistent polarized valence quadruple-zeta basis sets for the C and H atoms \cite{dunning:89}, augmented with corresponding F12 auxiliary density fitting basis sets (aug-cc-pVQZ). The H$_2$ geometry was fixed using the diatomic distance $r_0 = 1.4487$ bohr, which corresponds to the average value for the ground vibrational state of H$_2$.  The diabatization of the two $A''$ states was performed by a quasi-diabatization algorithm implemented in the MOLPRO program \cite{MOLPRO_brief}. A more detailed description of the PES calculations can be found elsewhere \cite{Klos2018}.

The QM CC calculations of ICSs and DCSs were performed using the HIBRIDON package \cite{2023CoPhC.28908761A}. In these calculations, the asymptotic experimental spin-orbit splitting of C($^3P$) ($A_{SO}=\Delta_{j=1}=16.41671$ cm$^{-1}$ and $\Delta_{j=2}=43.41350$ cm$^{-1}$) \cite{Harris2017} was used. The close-coupling equations were propagated from $R = 1.0$ to 80 bohr using the hybrid Alexander-Manolopoulos propagator \cite{Alexander:JCP:1987}, with $R$ denoting the C-H$_2$ inter-particle distance. The reduced mass of the C-H$_2$ complex is $\mu_{r} = 1.72577$ u. The cross sections were checked for convergence with respect to the inclusion of a sufficient number of partial waves and energetically closed channels. The H$_2$ basis included all levels with a rotational quantum number $j \leq 6$ belonging to the ground vibrational state manifold. At $E_{coll} \sim 150$ cm$^{-1}$ the contributions of the first 41 partial waves were included in the calculations. State-to-state (de-)excitation cross sections were obtained for transitions between all the fine structure levels of C($^3P_j$) over the collision energy range relevant to the experimental conditions (0 - 150 cm$^{-1}$) on a grid with a step of 0.1 cm$^{-1}$ (ICSs) or 0.2 cm$^{-1}$ (DCSs). The DCSs were computed on a grid of scattering angles spanning from $\degree{0}$ to $\degree{180}$ with a step of $\degree{1}$. The effective DCSs that were used as input for the image simulations were constructed from the computed DCSs by taking into account the experimental collision energy spreads as a Gaussian distribution. The 0.2 cm$^{-1}$ grid provides adequate sampling of the DCSs over the Gaussian distributions, except at $E_{coll} = 0.5$~cm$^{-1}$ where the energy spread is somewhat undersampled.

\section*{Acknowledgments}
S.Y.T.v.d.M. acknowledges financial support from the European Research Council (Consolidator Grant FICOMOL, Grant Agreement No. 817947). F.L. acknowledges financial support from the European Research Council (Consolidator Grant COLLEXISM, Grant Agreement No. 811363) and from Rennes Metropole. J.O. acknowledges support from the European Unions Marie Sk{\l}odowska-Curie Actions (Grant agreement No. 886046). X.-D.W. acknowledges support from the European Unions Marie Sk{\l}odowska-Curie Actions (Grant agreement No. 889328). This work is part of the research program of the Dutch Research Council (NWO). We thank Stach Kuijpers for discussions on determining the effective collision angle. The expert technical support by Niek Janssen, Andr\'e van Roij, Edwin Sweers and Gerben Wulterkens is gratefully acknowledged.

\bibliography{string,2023_C-H2_Zeeman_adj}

\end{document}